\documentclass[fleqn,usenatbib]{mnras}

\usepackage{newtxtext,newtxmath}

\usepackage[T1]{fontenc}

\DeclareRobustCommand{\VAN}[3]{#2}
\let\VANthebibliography\thebibliography
\def\thebibliography{\DeclareRobustCommand{\VAN}[3]{##3}\VANthebibliography}

\usepackage{graphicx}	
\usepackage{amsmath}	

\def\P     {P$_{1400}$ }

\def\le    {L$_{\rm Edd}$ }

\def \arcsec      {\text{$^{\prime\prime}$}}

\def \mujybeam    {$\muup$Jy\,beam$^{-1}$}

\title[RAD@home discovery of ring structures including Odd Radio Circles]
{RAD@home discovery of extragalactic radio rings and odd radio circles: clues to their origins}

\author[Ananda Hota et al.]{Ananda Hota,$^{1,2,3}$\thanks{E-mail:hotaananda@gmail.com}
Pratik Dabhade,$^{4,3}$\thanks{pratik.dabhade@ncbj.gov.pl}, Prasun Machado,$^{5,3}$, 
Joydeep Das,$^{6,3}$
Aarti Muley,$^{5}$
and Arundhati Purohit$^{3}$
\\
$^{1}$ Centre for Excellence in Theoretical and Computational Science, University of Mumbai, Santacruz-East, Mumbai-400098, India\\
$^{2}$ UM-DAE Centre for Excellence in Basic Sciences, University of Mumbai, Santacruz-East, Mumbai-400098, India\\
$^{3}$ RAD@home Astronomy Collaboratory, Kharghar, Navi Mumbai, PIN 410210, India\\
$^{4}$ Astrophysics Division, National Centre for Nuclear Research, Pasteura 7, 02-093 Warsaw, Poland\\
$^{5}$ Department of Physics, SIES College of Arts, Science \& Commerce, University of Mumbai, Sion (W), Mumbai 400 022, India\\
$^{6}$ School of Physics, Indian Institute of Science Education and Research Thiruvananthapuram, Thiruvananthapuram, 695551, India}

\pubyear{2025}

\begin{document}
\label{firstpage}
\pagerange{\pageref{firstpage}--\pageref{lastpage}}
\maketitle
\begin{abstract}
We present three rare and striking extragalactic radio sources discovered through visual inspection of low-frequency continuum maps from LoTSS DR2 and TGSS by the RAD@home citizen-science collaboratory. The first, RAD J131346.9+500320, is the first clear Odd Radio Circle (ORC) identified in LoTSS. At $z_{\rm phot}\simeq0.94$, it hosts a pair of intersecting rings of $\sim$300 kpc diameter, embedded in diffuse emission extending over $\sim$800 kpc, making it both the most distant and most powerful ORC reported to date. Its steep spectrum ($\alpha_{54}^{144}=1.22\pm0.15$) points to a relic synchrotron origin. The second object, RAD J122622.6$+$640622, is a $\sim$865 kpc giant radio galaxy whose southern jet is abruptly deflected, inflating a 100 kpc limb-brightened ring, while the northern jet terminates in a compact hotspot-like feature. The third, RAD J142004.0+621715 (440 kpc), shows a comparable ring at the end of its northern filamentary jet, along with a secondary filament parallel to its southern jet. All three systems lie in $\sim10^{14}M_\odot$ clusters or group-scale haloes, suggesting that environmental density gradients and possible jet–galaxy interactions play a central role in shaping these ring morphologies. These discoveries expand the zoo of extragalactic radio morphologies, highlight the diversity of pathways that can generate ring-like synchrotron structures, and demonstrate the continuing importance of human pattern recognition in identifying rare sources that escape current automated pipelines.
\end{abstract}



\section{Introduction}\label{sec:1_intro}
Since the discovery that Cygnus~A is a double radio source associated with an optical galaxy \citep{1953Natur.172..996J}, bipolar jets have dominated our understanding of radio galaxies (RGs), giving rise to linear FR I and FR II morphologies, wide-angle tails, S and X shapes, double-double restarted sources, and even megaparsec scale RGs \citep[for reviews see;][]{2019ARA&A..57..467B,HardcastleCrostron20,2022saikiajet,GRGreview}. 

Continuum radio astronomy has been transformed in the past decade due to the modern and powerful telescopes like the Low Frequency Array (LoFAR), the upgraded Giant Meterwave Radio Telescope (uGMRT), the Karoo Array Telescope (MeerKAT), the Australian Square Kilometre Array Pathfinder (ASKAP), and the Karl G. Jansky Very Large Array (JVLA). They deliver arcsecond to sub–arcsecond images at micro-Janskys sensitivity.  All sky surveys from these telescopes, both old and new, like NRAO VLA Sky Survey \citep[NVSS;][]{nvss}, TIFR GMRT Sky Survey Alternative Data Release 1 \citep[TGSS\,ADR1; ][]{tgss_intema}, LOFAR Two-metre Sky Survey \citep[LoTSS;][]{lotssdr2}, Evolutionary Map of the Universe \citep[EMU;][]{Hopkins2025EMU} and Very Large Array Sky Survey \citep[VLASS;][]{vlass} image tens of millions of sources across a broad frequency range, from $\sim$150 MHz up to 3 GHz. In radio interferometric imaging, low- and high-resolution maps often serve complementary roles, with the former revealing diffuse, extended structures that may be missed in higher-resolution survey images.

Recent deep and wide-field radio surveys have begun to reveal morphologies that extend beyond the classical jet paradigm. Examples include large-scale kinks in jets \citep[e.g.][]{2022A&A...668A..64D}-overgrown instabilities that can bend otherwise linear jets over scales of up to $\sim$100 kpc, and collimated synchrotron threads (CSTs) \citep[CSTs;][]{Ramatsoku20}, narrow linear filaments extending up to $\sim$100 kpc, either linking the lobes or protruding beyond them.

Among the most striking new phenomena uncovered in recent wide-field radio surveys are the \textit{odd radio circles} (ORCs), faint and roughly circular rings of radio emission up to $\sim$500~kpc in diameter that surround apparently normal galaxies yet show no obvious jets or counterparts at other wavelengths. Since their initial discovery with ASKAP \citep{2021PASA...38....3N}, only about half a dozen convincing examples have been reported, including one with the GMRT and several more with MeerKAT \citep{2025MNRAS.537L..42N} and EMU \citep{GuptaORC2025,KoribalskiORC2025}. ORCs are detected almost exclusively in deep low-frequency radio surveys such as ASKAP, GMRT, MeerKAT and EMU.
ORCs are typically of very low surface brightness, at the level of only a few mJy, and present a spherical or ring-shaped morphology in the radio continuum. Their radio luminosities span $\sim10^{23}$ to $10^{24}$~W\,Hz$^{-1}$ at $\sim$1\,GHz. Many ORCs have a central galaxy that is plausibly the host, while in some cases, no clear optical counterpart has been identified, which adds to the mystery. Their radio spectra are often steep, consistent with an ageing synchrotron population. These properties do not match established classes such as supernova remnants, planetary nebulae or classical radio galaxies, and they have prompted the view that ORCs may be fossil synchrotron shells that have been re energised by shocks or superwinds, or by blast waves from powerful extragalactic events.

For classification purposes, a ring-shaped radio source is considered an extragalactic ORC if it cannot be explained as a Galactic supernova remnant and, using the redshift of its central galaxy, corresponds to a physical scale of several hundred kiloparsecs. ORCs may appear as single rings or, more rarely, as two intersecting rings. By contrast, a ring that is clearly part of a lobe or hotspot in a nearby radio galaxy is not classified as an ORC. The scarcity of ORCs and their steep spectra suggest a rare and possibly short-lived phase of radio galaxy or feedback activity, although their true origin remains uncertain.

Ring-shaped radio structures, however, were recognised long before the identification of ORCs. Some occur as part of the diffuse emission within radio galaxy lobes \citep[e.g. 3C~310;][]{3C310-ring-vanBreugel1984}, while others are entirely unrelated to radio jets, such as gravitational lens systems \citep[e.g. the Ooty Lens PKS~1830$-$211;][]{OotyLensRaoSubrahmanyan1988,LensJauncey1991}.
  
The complex and often subtle morphologies of these apparently rare objects make them challenging to detect automatically, as current machine-learning pipelines are still being refined to identify extended, low-surface-brightness radio structures with irregular geometries \citep[e.g.,][]{2023Ndungu}. In addition, radio interferometric imaging is inherently sensitive to angular scale, with different configurations filtering out emission on certain size scales, further complicating interpretation compared to optical imaging. These factors make the reliable identification of such sources particularly difficult in large surveys. Consequently, careful visual inspection, especially when incorporating heterogeneous, multi-frequency, and multi-wavelength datasets- remains an indispensable approach for recognising and classifying complex radio structures such as ORCs.

The RAD@home citizen–science collaboratory has systematically examined hundreds of fields from the TGSS, integrating complementary data from surveys such as NVSS, FIRST, and LoTSS. This approach, combining wide-field low-frequency imaging with higher-resolution/frequency and multi-wavelength counterparts, has led to the discovery of numerous unusual and morphologically complex radio sources \citep{Hota2022RAD12, 2024Hota, Apoorva2025}. In this brief report, we highlight three representative cases: (1) the most distant ORC, (2) a giant radio galaxy (GRG) hosting a single, $\sim$\,100~kpc scale ring structure within its lobe, and (3) a jet–ring system exhibiting twin filaments (see Tab.~\ref{tab:general}). Each of these objects offers unique observational constraints on proposed formation mechanisms for radio rings, providing new tests for scenarios involving jet deflection, backflow-induced vortices, and large-scale wind-driven bubbles. Together, they illustrate the diversity of ring-like structures in extragalactic radio sources and the value of coordinated visual inspection across multiple radio surveys.

In this paper, we adopt the flat $\Lambda$CDM cosmological model based on the Planck results \citep{Planck}, with parameters $H_0$ = 67.4 km s$^{-1}$ Mpc$^{-1}$ (Hubble-Lema\^itre constant), $\Omega_m$ = 0.315, and $\Omega_{\Lambda}$ = 0.685 . The following radio-spectral index convention is being adopted: \( S_{\nu} \propto \nu^{-\alpha} \), where \( S_{\nu} \) is the flux density at frequency \( \nu \) and \( \alpha \) is the spectral index. The coordinates in all images are in the J2000 system.

\section{Citizen science discovery process}\label{sec:discovery}
RAD@home\footnote{\url{https://www.radathomeindia.org}} is a citizen science research collaboratory founded in 2013 \citep{Hota2014} as a zero–funding, zero-infrastructure, inter-university network that links professional astronomers with trained volunteers across India \citep[for more details, see;][]{Hota16}.  Participants range from undergraduate and postgraduate science or engineering students to motivated citizens with university degrees in any science discipline. For many students who lack access to conventional internships, live weekend sessions (e-class) provide their first hands-on experience of astronomical research. Training begins with \emph{Daily Galaxy RGB Analysis}: participants use the web-based RAD@home RGB–maker tool\footnote{\url{https://www.radathomeindia.org/rgbmaker}} to overlay radio contours on multi-band (UV, optical, IR) images of NGC and 3CR galaxies, following the workflow described by \citet{2023IAUS..375...40K}.  Once proficient, participants are promoted to the `i-astronomer' group and learn to inspect FITS images from large radio surveys (e.g., TGSS, LoTSS and VLASS). Any non-standard source uncovered is posted to the collaboratory forum, discussed by peers and vetted by professional mentors.

Furthermore, promising objects are submitted for follow-up with the GMRT through the GOOD-RAC (GMRT Observation of Objects Discovered by RAD@home Astronomy Collaboratory) proposal series.  When feasible, week-long discovery camps at Indian institutes complement the online training.  Several discoveries have already been reported \citep[e.g.,][]{2024Hota}, for example, specifically RAD\,12, a radio AGN that inflates a large unipolar bubble onto its merging companion \citep{Hota2022RAD12}.
\section{Results}

\begin{table*}
    \centering 
    \caption{Basic Data on radio sources with rings found by RAD@home citizen science} 
    \label{tab:general} 
\begin{tabular}{|c|c|c|c|c|} 
        \hline 
        Radio Source name & Host galaxy name  & Host redshift & largest size of ring region & 144~MHz Power (W Hz$^{-1}$)\\ 
        \hline 
        \hline 
         RAD J131346.9+500320 & SDSS J131346.92+500319.3  &$z_{\rm phot} = 0.937\pm0.045$ & 800~kpc& 2.27 $\times$ 10$^{26}$ \\ 
        \hline 
        RAD J122622.6$+$640622 & 2MASX J12262260+6406222 & $z_{\rm spec}=0.11024 \pm 0.00002$ & 100~kpc & 2 $\times$ 10$^{25}$ \\ 
        \hline 
        RAD J142004.0+621715 & 2MASX J14200494+6217147 & $z_{\rm spec}=0.14140 \pm 0.00003$ & 64~kpc& 5.47 $\times$ 10$^{22}$\\ 
        \hline 
    \end{tabular} 
\end{table*}

\begin{figure*}
    \centering
     \includegraphics[scale=0.169]{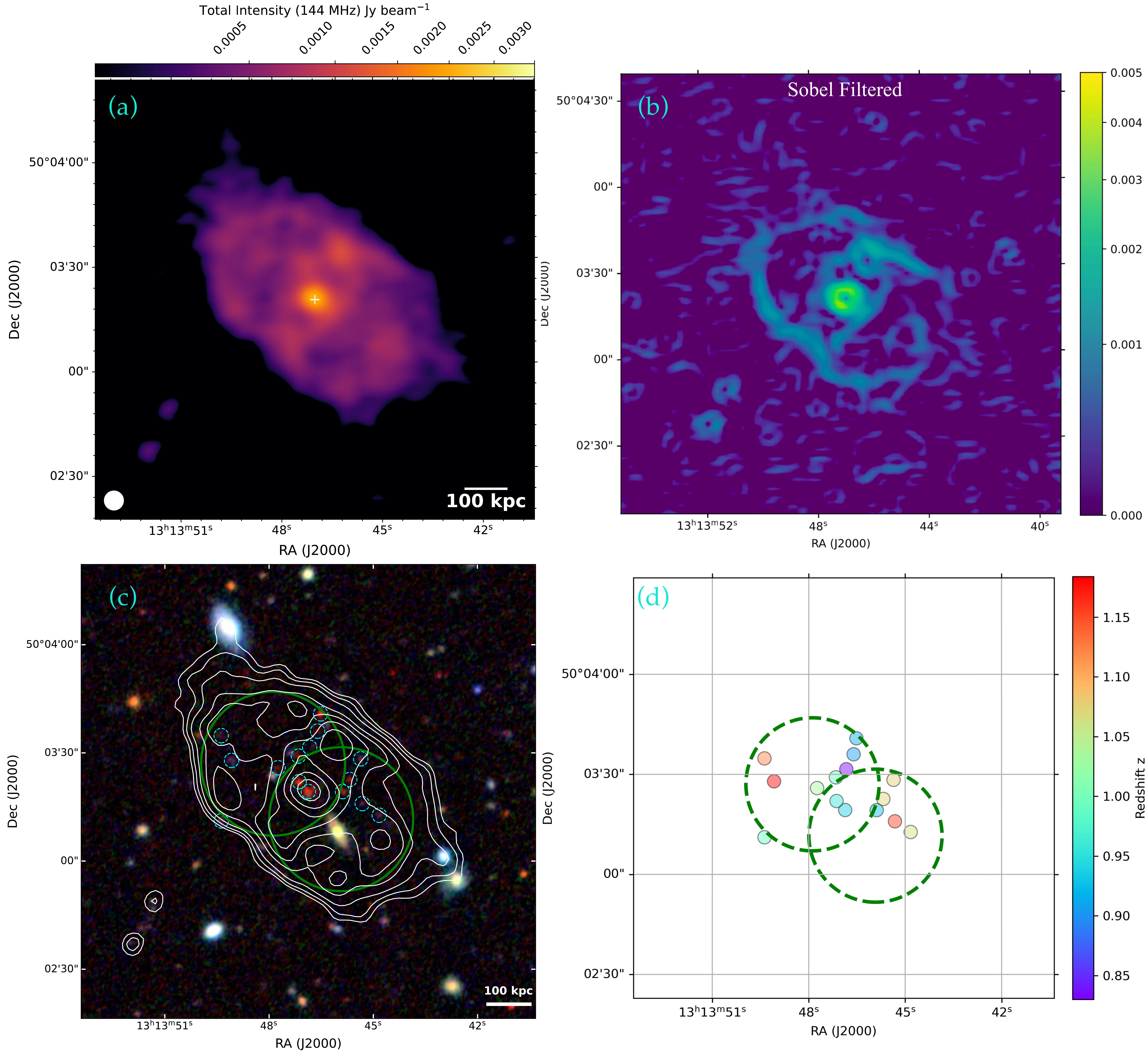}   
    \caption{(a):LoTSS 144~MHz radio image with 6\arcsec angular resolution of RAD J131346.9+500320, where only emission above $3\sigma$ is shown, and the $\sigma$ (rms noise) is $\sim$\,50 \mujybeam. The white cross shows the location of the object's host galaxy. (b): Edge-enhanced radio map of RAD J131346.9+500320 (Sec.~\ref{sec:orc}). The image was convolved with the standard $3\times3$ Sobel kernels in the horizontal ($G_x$) and vertical ($G_y$) directions, and the resulting pixel intensity is given by $\sqrt{G_x^2 + G_y^2}$, thereby highlighting sharp brightness gradients and filamentary structures. (c) LoTSS 144~MHz radio contours ($10^{-3} \times$ [0.15, 0.19, 0.26, 0.39, 0.49, 0.62, 0.81, 1.01, 1.41, 1.89, 2.54, 4.66, 8.63, 16.07, 21.96, 30.01] Jy~beam$^{-1}$) at $6^{\prime\prime}$ resolution in white, overlaid on the BASS optical image. Two green circles highlight the rings; cyan dashed circles mark galaxies at similar redshifts (for more, see Tab.~\ref{tab:radorcenv}). (d) Same as (c), but galaxies are colour-coded by redshift (see colour bar). The two large dashed green circles again indicate the rings.}
    \label{fig:ORC-LOTSS}
\end{figure*}

\begin{table}
 \caption{ List of galaxies shown in Fig.~\ref{fig:ORC-LOTSS}c \& d for which the redshift values are similar to that of RAD J131346.9+500320. The photometric redshifts are obtained from \citet{Duncan2022}. }
 \label{tab:radorcenv}
\begin{tabular}{lccc}
\hline
 N &        RA (deg) &      Dec (deg) & $\rm z_{phot} \pm z_{err}$ \\
\hline

 1 & 198.44528 & 50.05536 & $0.937 \pm 0.045$ \\
 2 & 198.44639 & 50.05611 & $0.956 \pm 0.056$ \\
 3 & 198.44119 & 50.05535 & $0.926 \pm 0.270$ \\
 4 & 198.44648 & 50.05809 & $0.990 \pm 0.081$ \\
 5 & 198.44892 & 50.05719 & $1.036 \pm 0.226$ \\
 6 & 198.44033 & 50.05629 & $1.073 \pm 0.102$ \\
 7 & 198.43883 & 50.05443 & $1.156 \pm 0.207$ \\
 8 & 198.44418 & 50.06001 & $0.905 \pm 0.088$ \\
 9 & 198.43901 & 50.05788 & $1.077 \pm 0.195$ \\
10 & 198.43679 & 50.05353 & $1.065 \pm 0.264$ \\
11 & 198.44382 & 50.06134 & $0.912 \pm 0.061$ \\
12 & 198.45450 & 50.05775 & $1.184 \pm 0.244$ \\
13 & 198.45574 & 50.05310 & $0.987 \pm 0.230$ \\
14 & 198.45574 & 50.05966 & $1.125 \pm 0.113$ \\
15 & 198.44513 & 50.05876 & $0.830 \pm 0.274$ \\
 \hline
 \end{tabular}
\end{table}

 \subsection{A symmetrical twin ring ORC at $z\sim$\,0.9 } \label{sec:orc}
The source was first identified during a RAD@home e-class on 2024 June 11, when two intersecting ring-like structures centred on a compact radio core were noticed in LoTSS 144-MHz data (Fig.~\ref{fig:ORC-LOTSS}a). The morphology exhibits all the defining characteristics of an ORC, and we designate the object RAD J131346.9+500320. This system had previously been classified\footnote{Although the source meets the conventional linear–size threshold for a giant radio galaxy (GRG), we do not designate it as such here because its ring–dominated morphology is atypical of the GRG class.} as a giant radio galaxy by \citet{2024Mostert}. To emphasise the sharp intensity gradients delineating the rings, we also present a Sobel-filtered image (Fig.~\ref{fig:ORC-LOTSS}b). An optical-radio map with LoTSS $6\arcsec$ resolution contours overlaid on optical at is shown in Fig.~\ref{fig:ORC-LOTSS}c, where green circles have been drawn to guide the eye to the double-ring structure. Mild brightness enhancements are visible at the ring intersections.  

The central compact radio peak ($1.75 \pm 0.18$ mJy) coincides with a faint optical galaxy (SDSS J131346.92+500319.3 or WISEA J131346.88+500320.0), more clearly seen in deeper Beijing–Arizona Sky Survey images \citep[BASS;][]{BASS}. This core is also detected in VLASS 3-GHz images (peak $= 0.75 \pm 0.13$~mJy\,beam$^{-1}$). Using these low and high frequency flux measurements of the central radio source, the spectral index of $\sim -$0.3 indicates a flat spectrum nature, typical of active radio galaxy cores. The central galaxy (hereafter host of the ORC) has a redshift of $z_{phot} = 0.937\pm0.045$ \citep{Duncan2022}.

No optical counterparts are found for secondary radio peaks along the rings, either in the foreground or background. However, the optical field reveals numerous galaxies within the ORC structure. Those with redshifts similar to the host ($z \sim 0.9$, within $\pm 1\sigma$) are marked with cyan circles (see Fig.~\ref{fig:ORC-LOTSS}c and Tab.~\ref{tab:radorcenv}), suggesting the presence of a galaxy group or poor cluster at $z \sim 0.9$. Two other systems are present but clearly unrelated: a faint galaxy at $z_{\rm phot} = 0.419 \pm 0.128$ located near the centre of the north-eastern ring, and an edge-on disc galaxy at $z_{\rm phot} = 0.297 \pm 0.020$ projected onto the south-western ring. Both lack the red colour characteristic of the $z \sim 0.9$ group, are unmarked by cyan circles, and show no detectable radio emission, confirming that they are unrelated to the ORC and do not affect its flux density measurements. Fig.~\ref{fig:ORC-LOTSS}d presents the spatial distribution of galaxies in the field, colour-coded by redshift, further illustrating the group/cluster environment associated with the ORC host.  

The twin rings, along with some diffuse emission, have a total angular extent of $\sim$\,100\arcsec, which corresponds to a projected linear size of about 800~kpc as measured from the LoTSS 6\arcsec angular resolution data. This value is comparable to the sizes of giant radio galaxies. Interestingly, the diffuse emission beyond the double-ring is aligned with the source's major axis, and that is also aligned with a small jet-like southward extension of the central radio core. 
Each of the rings is about 40\arcsec or 300 kpc in diameter. The complete structure has an integrated flux density of $43.2\pm4.1$~mJy, with the central core contributing $1.75\pm0.18$~mJy.  The ORC is only marginally detected in NVSS (3$\sigma$), with an integrated flux density of $2.8 \pm 0.6$~mJy. However, it is fully detected in the 54~MHz LOFAR LBA Sky Survey \citep[LoLSS;][]{lolssdr1} with a 15\arcsec\ beam, although at this resolution its detailed ring morphology is not resolved. From the LoLSS maps, our measurement—obtained by integrating the emission above the $3\sigma$ level within a polygon aperture, yields an integrated flux density of $143 \pm 16$~mJy. Using the integrated flux densities from LoLSS, LoTSS, and NVSS, we estimate the spectral index to be $\alpha_{54}^{144} = 1.22 \pm 0.15$ and $\alpha_{144}^{1400} = 1.20 \pm 0.10$. 

The consistently steep values over this wide frequency range support the interpretation of the ring as aged synchrotron plasma, characteristic of relic emission rather than ongoing jet activity. 

Using the integrated flux density of $43.2$~mJy at $144$~MHz and a spectral index of $\alpha_{144}^{1400} = 1.20$, we estimate the radio luminosity of the ORC to be $2.27 \times 10^{26}$~W~Hz$^{-1}$. This places it among the most powerful radio sources of its class, with a luminosity that is nearly two orders of magnitude higher than that of other known ORCs, which typically lie in the range $10^{23}$--$10^{24}$~W~Hz$^{-1}$ \citep[e.g., at 944~MHz;][]{GuptaORC2025}.

If the two rings were viewed face-on and well-aligned, they would appear as a single circular structure (O). In contrast, if they are seen exactly edge-on, it is unclear whether they would appear as two linear sources with the host galaxy in the centre, or still as two distinct rings. If these structures are in fact hollow spheres or shells, they may resemble two rings from most viewing angles. 
Since the three-dimensional geometry of ORCs remains unknown due to the uncertainty about their origin, the current view of intersecting circles likely results from an intermediate inclination. A similar concentric ring-like structure or two rings has also been noted in ORC J2103-6200 \citep{2022Norris}. The detection of these faint-ring structures of RAD J131346.9+500320 was made possible by the superior surface brightness sensitivity and adequate angular resolution of LoTSS. Among the close to a dozen ORCs discovered so far, RAD J131346.9+500320, with its  high redshift ($z \sim$\,0.9) and high luminosity, stands out as the most distant and most powerful ORC identified to date, and notably the first such discovery made using LoTSS data.

\subsection {Ring at the end of a diverted backflow in a giant radio galaxy}
\begin{figure*}
    \includegraphics[scale=0.185]{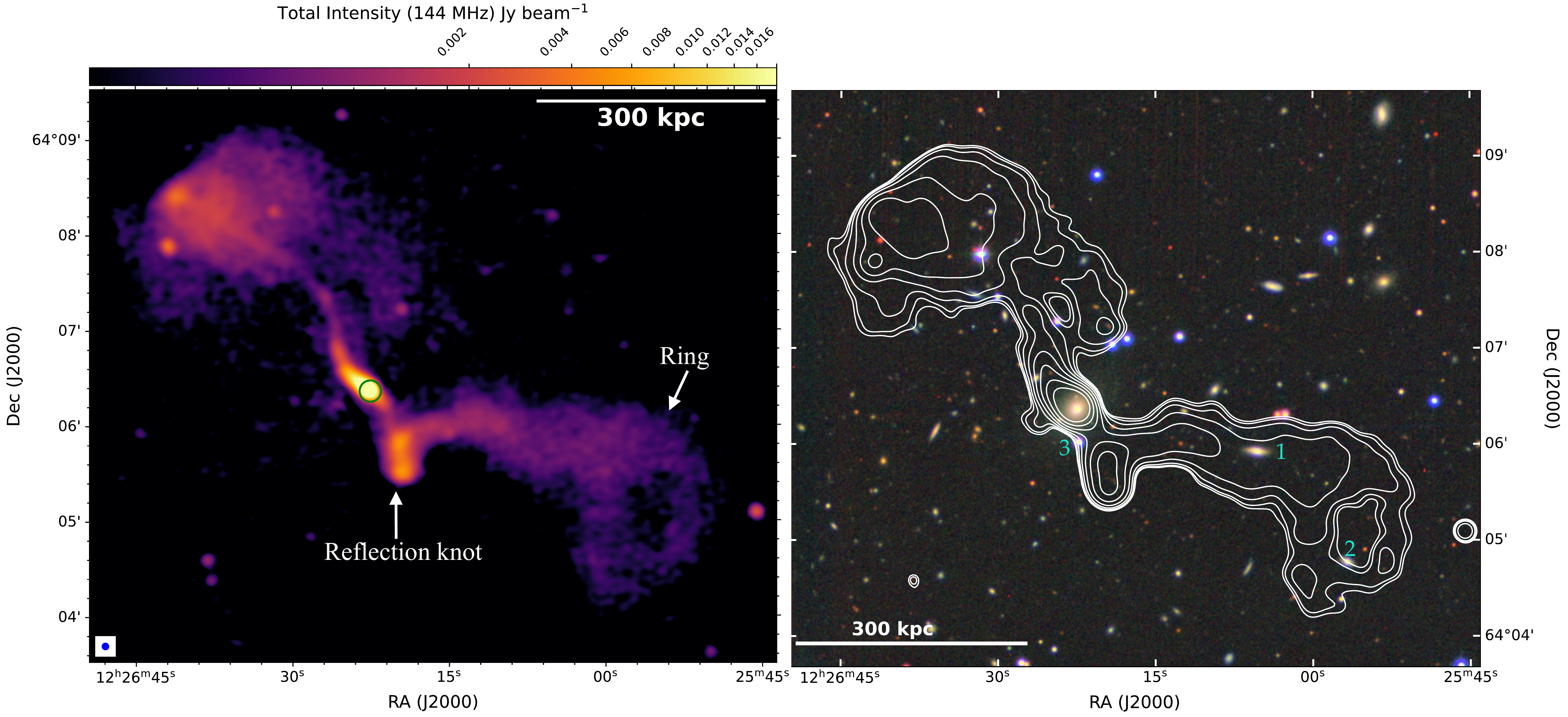}
    \caption{\textit{Left:}LoTSS 6\arcsec~ radio image of  RAD~J122622.6+640622 showing emission above $3\sigma$, where the $\sigma$ (rms noise) is $\sim$\,50 \mujybeam. The green circle shows the location of host BCG. \textit{Right:} LoTSS 6\arcsec resolution radio contours in white (levels = $10^{-3} \times [0.24,\,0.27,\,0.33,\,0.45,\,0.72,\,1.30,\,2.55,\,5.25,\,23.57]$~Jy~beam$^{-1}$) overlaid on the BASS RGB optical image. Galaxy~1 is a neighbouring system at a similar redshift as the host BCG, while Galaxies~2 and~3 are foreground objects unrelated to the source.}
\label{fig:rad-ring}
\end{figure*}

TGSSADR~J122622.6$+$640622 was first identified during an online e-class on 2022 August 13, when its TGSS morphology was recognised as the compact nucleus of a much larger, structurally asymmetric, double-lobed radio source visible in NVSS. The extended lobes are absent in both TGSS and FIRST, but a higher-quality LoTSS image (Fig.~\ref{fig:rad-ring}, left panel) reveals the complex nature of the source. As illustrated in the overlay of LoTSS radio contours on a BASS optical image (Fig.~\ref{fig:rad-ring}, right panel), bright bipolar well-collimated jets emerge from the host galaxy 2MASX~J12262260$+$6406222 or WISEA~J122622.51$+$640622.1 at a $z_{\rm spec}=0.11024 \pm 0.00002$ \citep{2012sdss}.  This galaxy is the brightest cluster member of J122622.5+640622 galaxy cluster \citep{2024WHDESI} and lies near a cosmic-web filament identified by \citet{Tempel2014}.  Close to the nucleus, the bipolar jets' brightness profile indicates an FR-I-type morphology. The north-eastern (NE) jet is slightly brighter than the south-western (SW) jet, suggesting the NE jet may be approaching the observer. The NE jet gradually fades before feeding a lobe that terminates in a mildly bright hotspot, echoing the behaviour of an FR-II. The S-shaped swing/precession of the NE-jet is also clearly seen. Similar to the NE, the SW jet is also aligned roughly with the minor axis of the host galaxy. On the southern side, the flow is deflected at a distinct \emph{reflection knot}, beyond which the plasma bends sharply westward, forms a diffuse plume, and ultimately traces a ring-like structure about $\sim 100$~kpc in diameter. The overall radio structure extends over $415\arcsec$ on the sky, corresponding to a projected linear size of 865~kpc, thereby qualifying it as a giant radio galaxy (GRG). The integrated flux density measured from the LoTSS map is 660$\pm$70 mJy, which yields an integrated 144~MHz radio power of $2 \times 10^{25}$~W\,Hz$^{-1}$. As the ring is unresolved in NVSS and a reliable flux-density estimate separating it from the western lobe could not be obtained, we did not attempt a spectral index calculation. This source was included in the GRG catalogue of \citet{2023OeiGRGs}, but to our knowledge it has not been studied in detail. We designate this remarkable system, comprising a giant radio galaxy with a $\sim 100$~kpc ring feature, as RAD~J122622.6$+$640622.

As seen in the radio--optical overlay, neither the double-peaked \emph{reflection knot} nor any other radio peak in the south-western lobe is associated with an optical counterpart. To investigate the origin of the complex morphology, we examined three nearby galaxies (labelled 1, 2, and 3 in Fig.~\ref{fig:rad-ring}, right panel).  
Galaxy~1 is an edge-on lenticular system located west of the host galaxy, with a photometric redshift of $z_{\rm phot}=0.106 \pm 0.006$, consistent with being a neighbour of the host. Interestingly, the westward radio plume appears to begin diverging and fading in brightness close to the position of this galaxy, raising the possibility of an interaction. Galaxy~2, another lenticular system, lies on the southern half of the ring and has a $z_{\rm phot}=0.065 \pm 0.007$. This redshift clearly places it in the foreground, ruling out any physical connection to the cluster or the lobe. A third system (Galaxy~3), located north-east of the reflection knot and to the south of the host galaxy, has $z_{\rm phot}=0.062 \pm 0.025$, again identifying it as a foreground galaxy.  
The overall radio structure is thus unusual: the source combines FR-I-like inner jets with an FR-II-like northern hotspot, while also exhibiting a peculiar reflection knot and a prominent southern ring. This hybrid morphology resists classification within the traditional Fanaroff–Riley scheme. In the absence of the reflection knot, the system might resemble a wide-angle tail radio galaxy; however, the reflection knot itself is highly puzzling. A slow deflection occurring along the line of sight can be seen as a sharp turn, in projection. Its presence strongly suggests a complex interaction, possibly involving the neighbouring galaxy (Galaxy~1) and/or dynamical effects from the intra-cluster medium.

The southern ring of RAD J122622.6+640622 closely mirrors the counter-jet `swirl' in NGC 7016 \citep{WorrallBirkinshaw2014}. In both cases, a jet is sharply deflected, its back-flow turns into a closed limb-brightened loop, and the jet on the opposite side remains almost straight. The NGC 7016 loop, $\sim$\,90~kpc from the core, fills a low-pressure X-ray cavity that lets buoyant plasma roll into a torus; a similar process is possible in RAD J122622.6+640622, where the jet bends at a reflection knot and expands into a large ring. 

The NGC 7016 swirl sits within the Abell 3744 cluster core, while the RAD J122622.6+640622 ring lies close to the edge of the $R_{500}\simeq590$\,kpc of its $M_{500}=0.8\times10^{14}\,M_\odot$ cluster, showing that back-flow structures can remain coherent out to the virial boundary. In NGC~7016, the rim coincides with a boundary between hot and cool gas; an analogous sliding layer may account for the sharp brightness jump and intact northern rim of the ring in RAD J122622.6+640622. Additionally, the observed ring need not be a closed loop but a helical structure of the jet/plume seen in projection as a ring. Future polarisation data may help clarify the underlying 3D geometry of the structure.

Possible explanations for the complex morphology are:
\begin{itemize}
    \item Jet deflection at a reflection knot $\sim$\,100 kpc from the core – the southern jet bends sharply westward at a projected distance comparable to the extent of the dense, rapidly cooling gas in the cluster centre (cluster cooling radius). Such a deflection suggests the presence of a dense obstacle or a steep pressure gradient in the surrounding ICM.

    \item Back-flow vortex ring at the edge of the cluster’s virial region – after the bend, the plasma broadens into a plume that curls into a limb-brightened loop $\sim$\,110 kpc in diameter. With $M_{500} \sim 0.8\times10^{14} M_\odot$ and $R_{500}=590$ kpc, this ring lies close to $R_{500}$ and may represent a buoyant back-flow bubble that has risen to the edge of the cluster’s virialised volume.

    \item Environmental asymmetry – the northern jet propagates relatively unimpeded to a hotspot, whereas the southern flow is deflected within 100~kpc. The implied north-east to south-west ICM density contrast is consistent with cool-core clusters that host BCG-centred cold fronts or filaments.
\end{itemize}

A definitive test of the ring’s origin requires sensitive polarisation and rotation measure (RM) imaging with LOFAR or JVLA to trace magnetic-field geometry, together with multi-frequency radio observations to constrain electron ageing, and deep X-ray observations to search for a co-spatial X-ray cavity (if any). These multi-wavelength diagnostics will determine whether the structure is a buoyant vortex bubble or a lobe–driven back-flow confined by the surrounding ICM.


\subsection{Ring at the end of a filamentary jet}\label{sec:filamentring}

\begin{figure*}
    \includegraphics[scale=0.18]{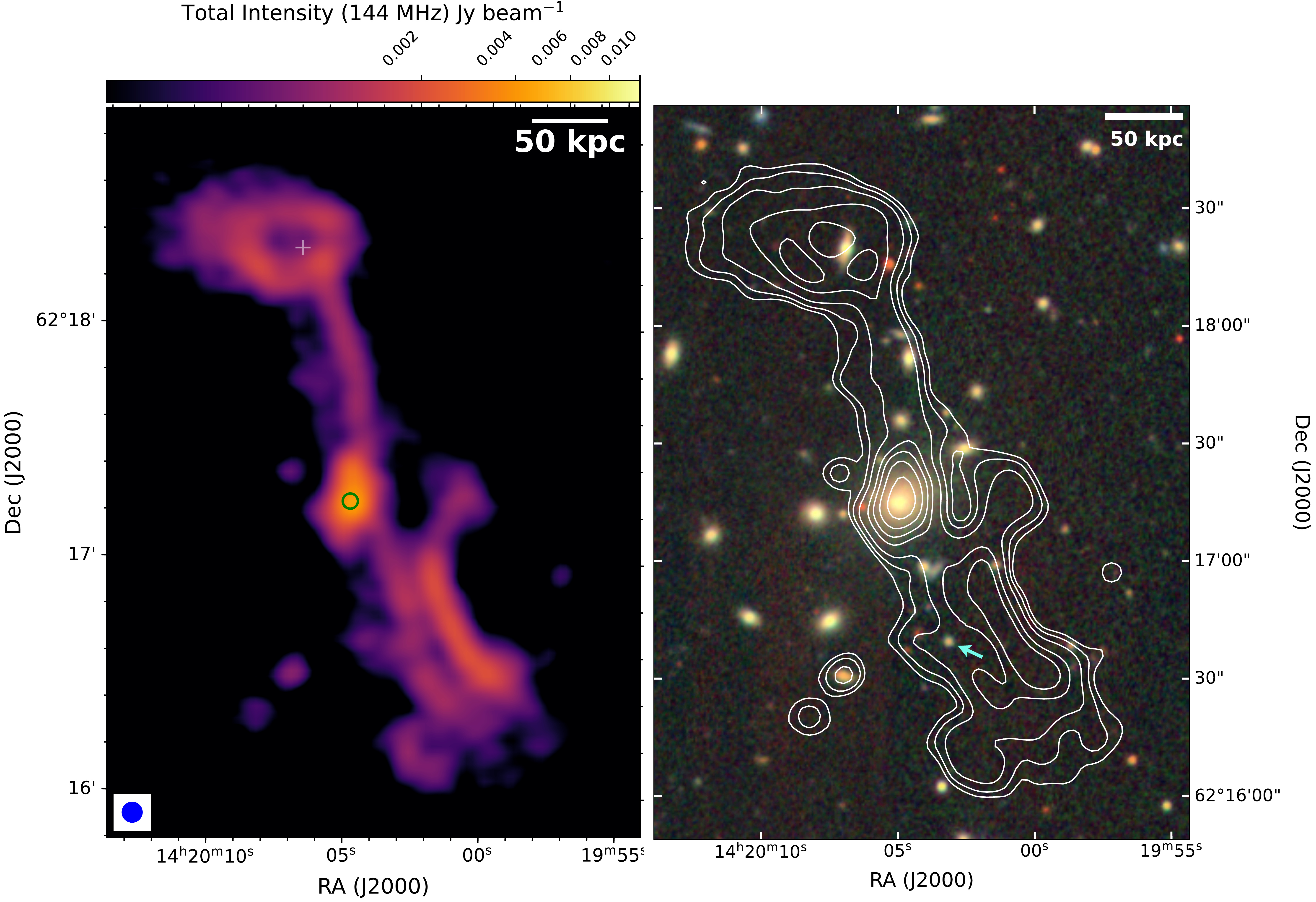}
    \caption{\textit{Left:}LoTSS 6\arcsec~ radio image of RAD J142004.0+621715 showing emission above $3\sigma$, where the $\sigma$ (rms noise) is $\sim$\,66 \mujybeam. The green circle shows the location of the host BCG. The white cross indicates the location of the edge-on disk galaxy as described in Sec.~\ref{sec:filamentring} and shown in more detail in Fig.~\ref{fig:ring2_north}. \textit{Right:} LoTSS $6^{\prime\prime}$ resolution radio contours in white 
(levels = $10^{-3} \times [0.19,\,0.30,\,0.49,\,0.91,\,1.29,\,1.86,\,2.73,\,4.01,\,5.95,\,8.85,\,13.22,\,19.77,\,29.60,\,44.37]$~Jy~beam$^{-1}$) 
overlaid on the BASS RGB optical image.
}
\label{fig:rad-ring2}
\end{figure*}

\begin{figure}
\centering
    \includegraphics[scale=0.38]{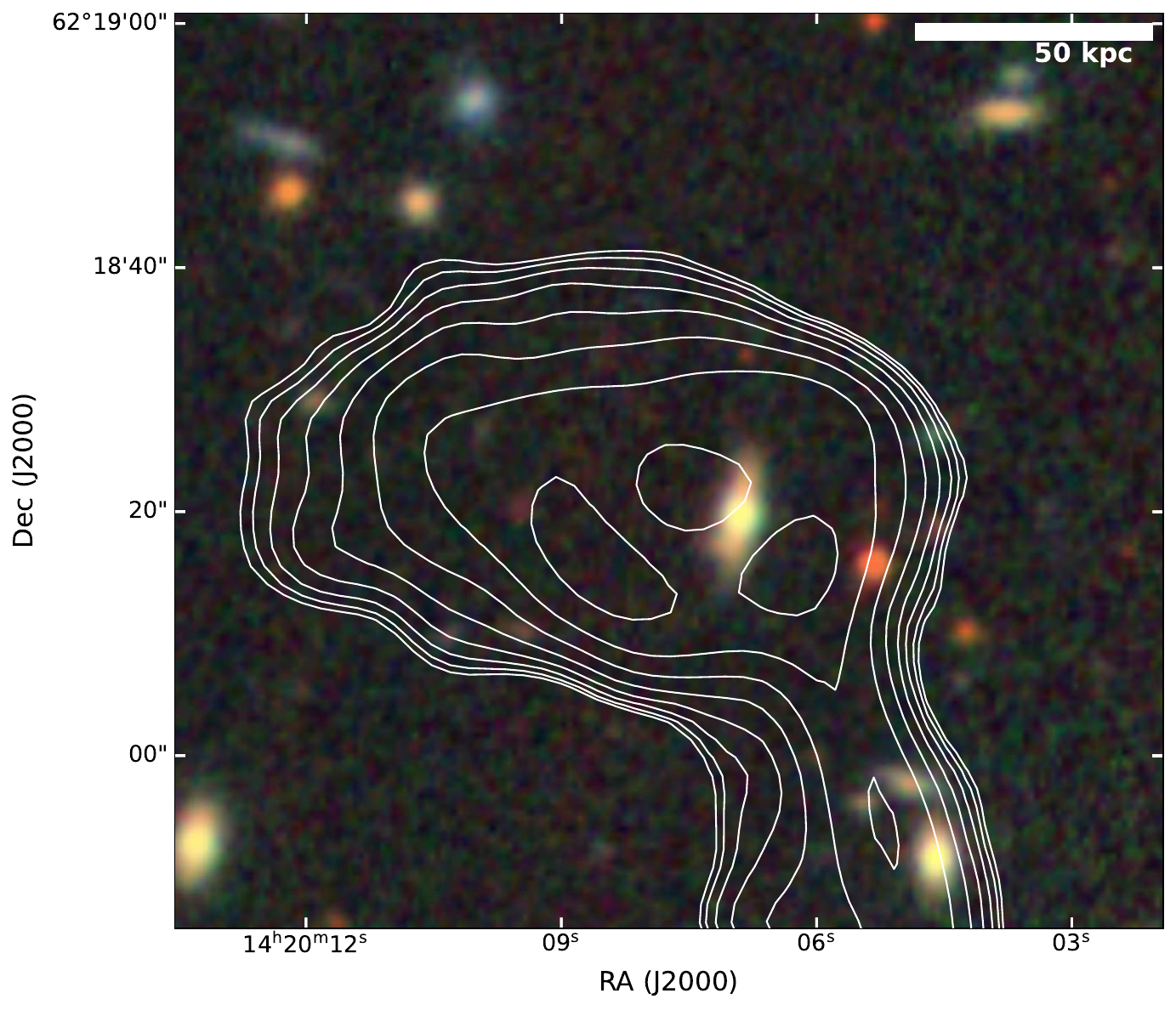}
    \caption{ Similar to the right sub-figure of Fig.~\ref{fig:rad-ring2} but zoomed in on the northern ring region to highlight structural details around the
    companion galaxy ($RA = 14^{\rm h}20^{\rm m}06\fs87$, $Dec =+62^{\circ}18'19\farcs1$).}
    \label{fig:ring2_north}

\end{figure}

\begin{figure}
    \centering
    \includegraphics[scale=0.45]{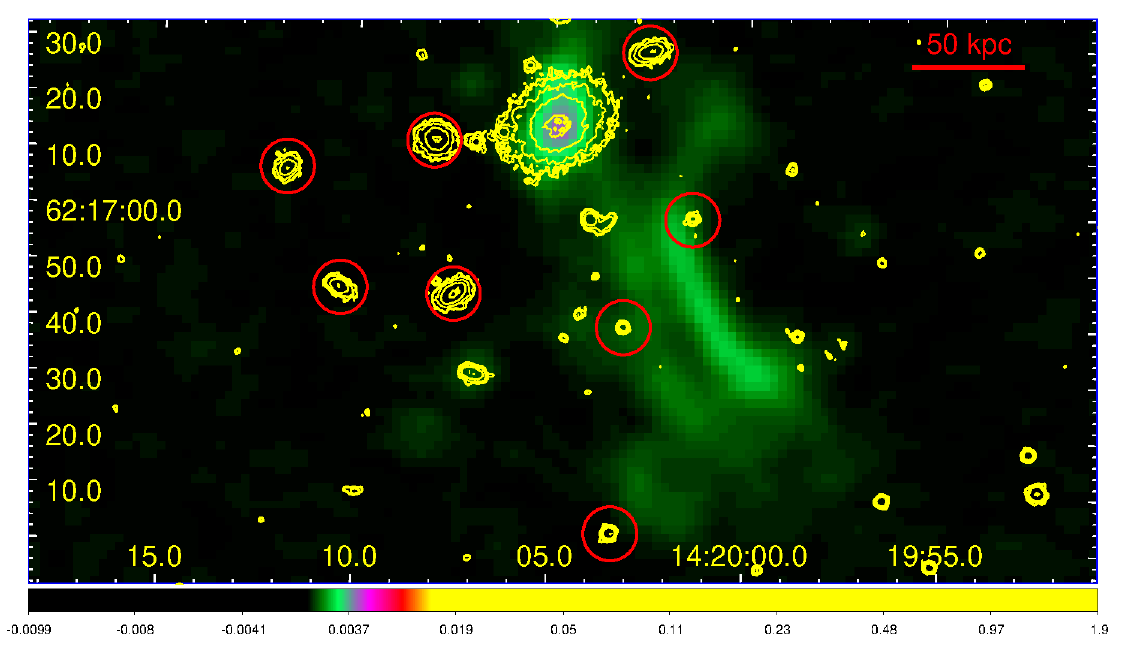}
    \caption{A false colour radio image from LoTSS with 6\arcsec angular resolution has been superposed with contours (yellow) of the optical image from BASS. Galaxies which are possible members of the cluster have been marked with a red circle.}
    \label{fig:Prasun-ring-southern-lobe}
\end{figure}

A non-standard filamentary radio galaxy, without any jet/plume, hotspot, or back flow structure, was first recognised in LoTSS\,DR2 data during the e-class of 2023 October 15. As seen in Fig.~\ref{fig:rad-ring2}, north of the nucleus (green circle), a loose/filamentary jet runs for $\sim 120$ kpc before broadening to form a limb–brightened radio ring ($\sim\,$64~kpc $\times$ 47~kpc). An overlay of the LoTSS radio image of this interesting radio source, RAD J142004.0+621715, on the optical image from BASS.  Fig.~\ref{fig:rad-ring2} shows the loose filamentary jet to emerge from a giant elliptical galaxy which is the brightest cluster galaxy (BCG), also known as MaxBCG J215.02067+62.28753 or 2MASX J14200494+6217147 or LEDA 2634038 ($R.A = 14^{\rm h}20^{\rm m}04\fs95$, $Dec = +62^{\circ}17'15\farcs1$; $z_{\rm spec}=0.14140 \pm 0.00003$; \citealt{2020SDSSEBOSS}). The host BCG belongs to a massive galaxy cluster - WHL~J142005.0+621715 \citep{whl12} of mass $M_{500}=1.63 \times 10^{14}\,M_\odot$ and virial radius of $R_{500}=825$ kpc. The whole source, north to south, extends $171\arcsec$ on the sky, giving a projected linear size of $\simeq 440$ kpc, with an integrated 144 MHz flux density of $104.2 \pm 10$ mJy or radio power of 5.47 $\times$ 10$^{22}$ W Hz$^{-1}$. To this integrated flux density, the ring on the north-east contributes $31.5 \pm 3$ mJy. The peaks on the ring are seen brighter than the surface brightness in the jet. There appears to be a bright spot at the location where the filamentary jet touches/enters the ring. (Fig.~\ref{fig:ring2_north}). 

The ring appears to move westward, as suggested by the compressed contours on the western edge and a tail or widely spaced contours seen on its eastern edge. The east-west elongation of the ring, which is nearly orthogonal to the jet, suggests its elongation is not due to the momentum of the jet, but some other force. An edge-on disk galaxy is observed to be on the western half of the ring ($RA = 14^{\rm h}20^{\rm m}06\fs87$, $Dec =+62^{\circ}18'19\farcs1$) and has a photometric redshift $z_{\rm phot}=0.1409$, consistent with cluster membership or it being a neighbouring galaxy. The east-west ring elongation is aligned with the minor axis of the disk galaxy. This disk galaxy has a colour $u-r = 3.3$, placing it firmly among early-type red quiescent systems despite its clear disc structure. A colour image (RGB composite of i-r-g band images) from BASS images reveals the disc to be redder than the bulge. This redness is probably due to the red-shifted H$\alpha$ line from the ionised gas entering the i-band filter. The southern half of the galaxy shows an extraplanar red finger-like feature extending from the mid-plane of the disk to the east. This faint feature is also visible in the Panoramic Survey Telescope and Rapid Response System (PanSTARRS) and Sloan Digital Sky Survey (SDSS) images, confirming its authenticity. The southern half of the red disk is also seen to be smaller than that in the north. This truncation of the southern disk is similar to the truncation of H\,{\sc i}  disks in ram-pressure-stripped galaxies. Targeted emission-line spectroscopy will be needed to decide whether this asymmetry in ionised gas is because of ram pressure stripping and/or excited by the surrounding radio plasma jet hitting from the southern side.

On the south of the core-galaxy region,  $\sim$\,160~kpc radio emission is observed, part of which appears to be a filamentary jet Fig.~\ref{fig:Prasun-ring-southern-lobe}. There seems to be a mismatch between the jet and the southern extension from the radio core. This is not the case on the northern side. West of this jet-like structure, a curved filamentary structure is observed. This C-shaped structure is as long as the jet, and its central region has surface brightness higher than the jet itself. C-shaped radio emission structures in clusters are usually head-tail radio galaxies or wide-angle tailed (WAT) radio galaxies, or peripheral radio relics related to shocks due to cluster mergers. The orientation of the C-shape with respect to the BCG clearly suggests it is not a cluster-relic. There is no optical galaxy seen on the C-shaped radio structure, ruling out the possibility that it is a head-tail or WAT. In the Fig. \ref{fig:Prasun-ring-southern-lobe}, contours from the BASS optical image have been plotted on the false colour image from LoTSS, where all the possible member galaxies have been marked with red circles. Along the jet structure, there is one optical galaxy ($RA = 14^{\rm h}20^{\rm m}03\fs13$, $Dec =+62^{\circ}16'39\farcs43$, marked with cyan arrow in Fig.~\ref{fig:rad-ring2}) having a photometric redshift of $z_{\rm phot}=0.159\pm0.034$ that is very likely a member of the cluster. It is intriguing that this galaxy is seen at the location where the linear jet structure shows a clear disruption.  The jet breaking at the location of the cluster member is also the closest to the mid-point of the C-shaped structure. There is a possibility that the pre-existing jet has been hit by the movement of the member galaxy and that has created the jet break and possibly part of this complex radio structure. It is unclear if the jet is physically connected to this C-shaped structure or if the proximity is only a projection effect.  Currently, we do not see any radio-optical clue to the origin of the C-shaped structure. It is extremely unusual that, despite an FR I-like jet being seen from this BCG and a total extent of $\sim$\,280~kpc, typical of FR I/FR II radio galaxies, neither the northern nor the southern side resembles standard FR I or FR II radio galaxies. This probably hints at a new kind of large filamentary radio galaxies where interaction of the dead/loose jet with neighbouring galaxies creates new structures that are not seen in standard radio galaxies.   

\vspace{-0.3cm}
\section{Discussion}
Radio rings are rare features in extragalactic astronomy, with only a handful of confirmed examples observed to date, spanning a variety of environments. Among them, ORCs remain the most enigmatic. In this paper, we report three new discoveries from the RAD@home citizen–science programme: (1) a twin, intersecting ring system in RAD J131346.9+500320, the most distant and most powerful ORC known to date ($z \sim \,0.94$); (2) a $\sim \,$100 kpc ring at the terminus of a diverted backflow in the giant ($\sim \,$865 kpc) radio galaxy RAD J122622.6+640622; and (3) a ring located at the end of a filamentary jet in RAD J142004+62171. Given the unresolved questions surrounding the origin of ORCs, these newly reported systems provide valuable observational constraints not only for models of ORC formation and evolution, but also for similar synchrotron rings and vortex structures containing relativistic magnetised plasma on scales from kiloparsecs to megaparsecs.

\subsection{Twin ORC from Speca with Superwind ?} 
\begin{figure}
    \centering
    \includegraphics[scale=0.45]{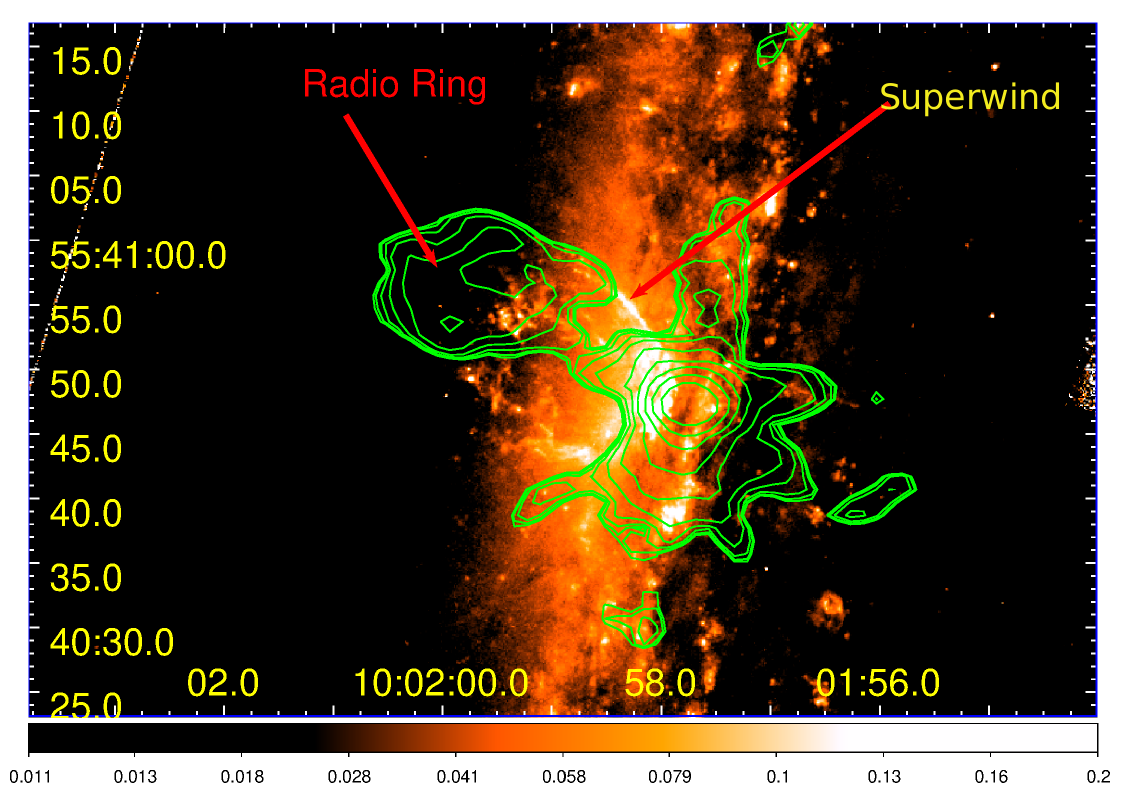}
    \caption{ NGC~3079: A false colour H$\alpha$+[N\,{\sc ii}] image from Hubble Space Telescope is superposed with contours (green) from radio emission from VLASS survey at 3 GHz. The ring can be seen at the top of the bowl-shaped filaments of ionised gas outflow (white) from the centre of the galaxy, seen as the peak of the radio emission.}
    \label{fig:NGC3079}
\end{figure}

ORCs with diameters of $\sim$\,100–500 kpc, steep radio spectra, and centrally located hosts are best understood as fossil radio shells re-energised by external/internal processes \citep[e.g.,][]{2024PASA...41...24S, DolagORC, IvlevaORC}. A large-scale shock induced by a galaxy–galaxy or black-hole merger, or a powerful superwind, could compress a dormant radio lobe and re-accelerate its particles to be visible as rings or broken shells \citep[for more, see;][]{KoribalskiORC2025}. The rarity of ORCs suggests that such geometries or coincidences in time are uncommon. Observational evidence from ORC 4, where a Keck spectrum revealed a $\sim$40~kpc ionised outflow with a wind-like velocity field \citep{2024Natur.625..459C}, demonstrates that fast thermal gas and re-brightened radio shells are likely causally connected and can coexist. A smaller-scale analogue is seen in the nearby Seyfert-starburst composite galaxy NGC~3079, where a $\sim$\,1\,kpc polarised radio ring is seen in the middle of a $\sim$\,10\,kpc wind-blown radio bubble \citep{2019ApJ...883..189S, NGC3079IrwinSaikia, CecilNGC3079}. As can be seen in Fig.~\ref{fig:NGC3079}, the radio ring (seen in the VLASS green contours) is located at the tip of the ionised gas outflow from the superwind (Hubble Space Telescope WFPC2 f658 filter image showing H$\alpha$ $+$ [N\,{\sc ii}] line emission). While the $\sim$\,1 kpc ring can be seen in the high resolution images of the nuclear lobe nearly 2 kpc away, the diffuse radio emission extends beyond that up to nearly 11 kpc from the stellar mid-plane  \citep[also see Fig. 3c and 8 in ][]{NGC3079IrwinSaikia}. Such expanding bubbles, given permissive environmental conditions, can scale up to ORC dimensions over a time period of a few hundred million years. To expand up to ORC-dimensions, hundreds of kpc, the radio jet is required to have that power which the Seyfert-starburst system NGC~3079 may not have. However, there is a growing class of Speca-like radio galaxies which are hosted by optically red disk/spiral galaxies and can grow in multiple episodes up to more than a Mpc scale \citep[Speca;][]{hotaspeca, LuisHoSpeca,Bagchi2025}. If a bipolar superwind, from the spiral host, starts after the radio lobes have reached a remnant/relic phase, a twin radio ring can possibly form and grow to large sizes. Although a single-circle ORC can be a shock-revived remnant radio lobe from any elliptical host galaxy, the possibility of twin rings seems easier with the presence of bipolar superwind from a Speca-like disk galaxy with large-scale relic radio lobes. RAD J131346.9+500320 seems to have 800~kpc diffuse emission within which 300~kpc rings are expanding along the major axis, which is also aligned with a jet-like feature from the central radio peak.     

\subsection{Radio rings from diverted backflow ?} 
The other environment where we find a radio ring is the GRG RAD~J122622.6$+$640622. In this source, as in NGC~7016, the radio jets undergo a sharp deflection, apparently caused by interaction with the circum-galactic or intra-cluster medium—no optical galaxy is seen at the location of the reflection knot. From this point, the jet backflow is redirected away from the main axis. Although no X-ray data are yet available for RAD~J122622.6$+$640622, the case of NGC~7016 provides a useful analogue: there, the backflow expands into a large cavity visible in the X-rays and subsequently forms a ring. A neighbouring galaxy, NGC~7018, also shows a filament of backflow plasma expanding towards the same cavity. By analogy, RAD~J122622.6$+$640622 may host a similar cavity, or alternatively, a steep density gradient in the surrounding ICM close to the host could be driving the deflection.

Notably, in both RAD~J122622.6$+$640622 and NGC~7016, the deflected radio structures extend roughly along the minor axis of the host galaxy. The combination of a reflection knot and a diverted backflow bears a strong resemblance to the secondary lobes or wings observed in X-shaped radio galaxies (XRGs), where such wings are also preferentially aligned with the host minor axis \citep{XRG-Capetti2002}. In RAD~J122622.6$+$640622, the westward plume of diverted backflow becomes faint and laterally broadened at the position of a nearby edge-on disk/lenticular galaxy. 

\vspace{-0.2cm}
\subsection{Radio rings from jet-galaxy interaction ?} 
Given the presence of a finger-like extra-planar feature and the possibility of disk-wide H$\alpha$ emission in the galaxy located at the centre of the radio ring in RAD~J142004.0$+$62171, one plausible explanation is the action of a galactic superwind. If such a wind exists, the combined radio–optical morphology can be understood as follows. As the galaxy moves westward and/or is impacted by a filamentary radio jet from the south-west, the non-thermal plasma on the western side would collide with the thermal outflow. Conversely, on the eastern side, the superwind could deflect the incoming plasma, producing a localised gap in the radio emission followed by a long downstream tail. This interaction could generate a limb-brightened ring structure.

An instructive analogy can be drawn with the interaction of the solar wind and the Earth’s magnetosphere. In that case, the solar wind compresses the magnetosphere on the Sun-facing side (the `dayside'), producing a bow-shaped front, while the deflected plasma forms a long magnetotail downstream. By extension, if a galactic superwind or gaseous halo encounters an impinging radio jet or relic plasma, the western side of the galaxy could produce a limb-brightened bow-like ring, while the displaced plasma on the opposite side could develop into a trailing tail. This framework naturally explains both the ring morphology and the associated downstream extension. The elevated surface brightness of the ring relative to the jet, together with the absence of detectable radio emission from the galaxy itself, further suggests localised re-acceleration at the interaction front.
        
A useful comparison can also be made with RAD~12 \citep{Hota2022RAD12}, where a radio jet striking a companion galaxy produces a depression in the synchrotron plasma and a mushroom-cloud-like radio feature. If such a system were observed at a later evolutionary stage, the depression might expand and the cloud evolve into a ring, as the plasma is unable to penetrate the dense interstellar medium of the massive companion. Future observations will be crucial in determining whether jet–galaxy interactions of this kind can indeed give rise to ring-like radio structures.

\subsection{Citizen Science and Machine learning} 
The first two of the three sources reported here had been catalogued earlier as giant radio galaxies, without noting the exotic ring structures embedded within them. Similarly, the third source not only has a ring but is too complex to interpret, even with the inclusion of optical colour images and redshift information of surrounding galaxies. In a recent study comparing automated source classifications in LOFAR imaging data and citizen science source classifications from the Radio Galaxy Zoo project, it was found that human performance is better for extended sources  \citep[Fig. 8 in][]{HardcastleMachineVsHuman2023}. Machine-learning pipelines have already become indispensable for sifting the millions of sources delivered by surveys such as LoTSS, EMU, VLASS and, soon, the SKA Observatory. They handle the bulk morphologies very efficiently, but inherently struggle with classes that are under-represented in the training sets.  For example, the automated search of \citet{2024Mostert} returned thousands of candidate GRGs, yet the intersecting ORC presented in this paper is listed there as an elongated double, where the algorithm overlooked the ring geometry that the eyes of a trained citizen scientist recognised at once. Such oversights are typical whenever the training set lacks true oddities.

Citizen-science programmes like RAD@home are designed to fill this gap. Volunteers, after structured training, impose no pre-defined template on the data and routinely flag the unexpected: ORCs, radio-lobe with rings, extreme kinks, collimated synchrotron threads, jet-galaxy interactions, etc. \citep[e.g.][]{2024Hota, Apoorva2025}. Their discoveries should then feed back into improved training sets for the next generation of neural networks. In the era of petabyte survey archives, progress will come from this partnership: algorithms for speed and completeness, human insight for novelty and context, rather than from either approach in isolation.

\vspace{-0.7cm}
\section*{Acknowledgements}
\begin{small}
We thank the Scientific Editor and the referee for their constructive comments that helped improve this manuscript. AH acknowledges the University Grants Commission (UGC, Ministry of Education, Govt. of India) for his monthly salary grants since June 2014. This research has made use of the VizieR catalogue tool, CDS, Strasbourg, France \citep{vizier}.  LOFAR data products were provided by the LOFAR Surveys Key Science Project (LSKSP; \url{https://lofar-surveys.org/}) and were derived from observations with the International LOFAR Telescope (ILT). LOFAR \citep{vanHaarlem2013} is the Low Frequency Array designed and constructed by ASTRON. It has observing, data processing, and data storage facilities in several countries, which are owned by various parties (each with its own funding sources), and which are collectively operated by the ILT foundation under a joint scientific policy. The efforts of the LSKSP have benefited from funding from the European Research Council, NOVA, NWO, CNRS-INSU, the SURF Co-operative, the UK Science and Technology Funding Council and the J\"{u}lich Supercomputing Centre. We acknowledge the use of the DESI Legacy Imaging Surveys (\url{https://www.legacysurvey.org/acknowledgment/}). For citizen science research and training, RAD@home has extensively used GMRT data from TGSS survey (both DR5 (P.I. Late Dr. Sandeep Sirothia) and ADR1 (P.I. Dr. Huib T. Intema)). We thank the staff of the GMRT who made these observations possible. GMRT is run by the National Centre for Radio Astrophysics of the Tata Institute of Fundamental Research.  

\end{small}
\vspace{-0.7cm}
 \section*{Data Availability}
 \begin{small}
This research has made use of all publicly available astronomy data. LoTSS and BASS data can be obtained via \url{https://lofar-surveys.org/releases.html} and \url{https://www.legacysurvey.org/bass/} or \url{https://casdc.china-vo.org/archive/BASS/DR3/}, respectively.  
\end{small}

\vspace{-0.7cm}
\bibliographystyle{mnras}
\bibliography{RAD-ORC-accepted} 




\bsp	
\label{lastpage}
\end{document}